\documentclass[sigconf]{acmart}
\usepackage{multirow}
\usepackage{subfigure}
\usepackage{blindtext}
\usepackage{enumitem}
\usepackage{algorithm}  
\usepackage{algpseudocode}  
\usepackage{amsmath}  
\usepackage{hyperref}
\usepackage{flushend}
\usepackage{balance}
\usepackage{mathtools}
\usepackage{bbm}
\DeclareMathOperator*{\maximize}{\text{\fontfamily{pcr}\selectfont maximize}}

\AtBeginDocument{%
\providecommand\BibTeX{{%
  \normalfont B\kern-0.5em{\scshape i\kern-0.25em b}\kern-0.8em\TeX}}}

\copyrightyear{2024}
\acmYear{2024}
\setcopyright{rightsretained}
\acmConference[SIGIR '24] {Proceedings of the 47th International ACM SIGIR Conference on Research and Development in Information Retrieval}{July 14--18, 2024}{Washington, DC, USA.}
\acmBooktitle{Proceedings of the 47th International ACM SIGIR Conference on Research and Development in Information Retrieval (SIGIR '24), July 14--18, 2024, Washington, DC, USA}
\acmISBN{979-8-4007-0431-4/24/07}
  \acmDOI{10.1145/3626772.3657774}

\begin{document}

\title{Deep Automated Mechanism Design for Integrating Ad Auction and Allocation in Feed}

\author{Xuejian Li}
\authornotemark[1]
\affiliation{%
 \institution{Meituan}
 \city{Beijing}
 \country{China}
}
\email{lixuejian03@meituan.com}

\author{Ze Wang}
\authornote{Corresponding authors.}
\affiliation{%
 \institution{Meituan}
 \city{Beijing}
 \country{China}
}
\email{wangze18@meituan.com}

\author{Bingqi Zhu}
\affiliation{%
 \institution{Meituan}
 \city{Beijing}
 \country{China}
}
\email{zhubingqi@meituan.com}

\author{Fei He}
\affiliation{%
 \institution{Meituan}
 \city{Beijing}
 \country{China}
}
\email{hefei11@meituan.com}

\author{Yongkang Wang}
\affiliation{%
 \institution{Meituan}
 \city{Beijing}
 \country{China}
}
\email{wangyongkang03@meituan.com}

\author{Xingxing Wang}
\affiliation{%
 \institution{Meituan}
 \city{Beijing}
 \country{China}
}
\email{wangxingxing04@meituan.com}

\renewcommand{\shortauthors}{Xuejian Li, et al.}

\begin{abstract}
E-commerce platforms usually present an ordered list, mixed with several organic items and an advertisement, in response to each user's page view request. This list, the outcome of ad auction and allocation processes, directly impacts the platform's ad revenue and gross merchandise volume (GMV). Specifically, the ad auction determines which ad is displayed and the corresponding payment, while the ad allocation decides the display positions of the advertisement and organic items. The prevalent methods of segregating the ad auction and allocation into two distinct stages face two problems: 1) Ad auction does not consider externalities, such as the influence of actual display position and context on ad Click-Through Rate (CTR); 2) The ad allocation, which utilizes the auction-winning ad's payment to determine the display position dynamically, fails to maintain incentive compatibility (IC) for the advertisement. For instance, in the auction stage employing the traditional Generalized Second Price (GSP) , even if the winning ad increases its bid, its payment remains unchanged. This implies that the advertisement cannot secure a better position and thus loses the opportunity to achieve higher utility in the subsequent ad allocation stage. Previous research often focused on one of the two stages, neglecting the two-stage problem, which may result in suboptimal outcomes.

Therefore, this paper proposes a deep automated mechanism that integrates ad auction and allocation, ensuring both IC and Individual Rationality (IR) in the presence of externalities while maximizing revenue and GMV. The mechanism takes candidate ads and the ordered list of organic items as input. For each candidate ad, several candidate allocations are generated by inserting the ad in different positions of the ordered list of organic items. For each candidate allocation, a list-wise model takes the entire allocation as input and outputs the predicted result for each ad and organic item to model the global externalities. Finally, an automated auction mechanism, modeled by deep neural networks, is executed to select the optimal allocation. Consequently, this mechanism simultaneously decides the ranking, payment, and display position of the ad. Furthermore, the proposed mechanism results in higher revenue and GMV than state-of-the-art baselines in offline experiments and online A/B tests. 

\end{abstract}


\maketitle

\section{Introduction}
\label{sec:1}
\begin{figure}[tb]
  \centering
  \includegraphics[width=0.7\linewidth]{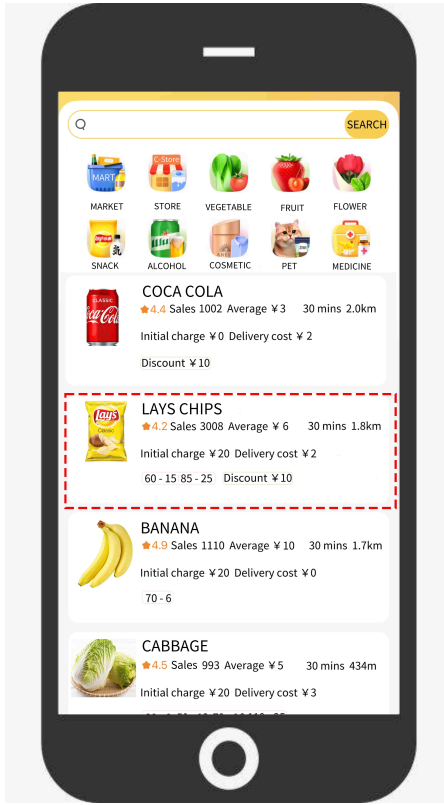}
  \caption{
    On Meituan retail delivery platform, the ad 'LAYS CHIPS' in the feed is presented to user along with external organic items. Whether a user clicks on the ad is easily impacted by both the position and the context of the ad.
  }
  \label{fig:fig_intro}
\end{figure}

In many e-commerce feeds, an ordered list, mixed with several organic items and a pay-per-click ad, is displayed for each page view request of users \cite{yan2020ads, zhang2018whole}, as is shown in Figure \ref{fig:fig_intro}. If users click on the ad, the platform will charge the corresponding advertiser, which is a key source of revenue for the platform \cite{qin2015sponsored, edelman2007internet}. Meanwhile, if users purchase the product in organic items or ad, the platform's gross merchandise volume (GMV) will increase. In practice, this list, produced by ad auction and allocation processes, directly determines the platform's revenue and GMV. More specifically, the ad auction determines which ad is displayed and the payment of the ad  \cite{edelman2007internet, varian2007position}, while the ad allocation determines the display positions of the ad and organic items in feeds \cite{zhang2018whole, liao2021cross}. Ad auction and allocation are interactive \cite{metrikov2014whole}. Hence, mechanism design for ad auction and allocation to maximize platform revenue and GMV has become a very meaningful and challenging problem.

The widely used method is to divide ad auction and allocation into two stages.  Initially, in the ad auction, such as the classic GSP \cite{varian2007position}, candidate ads are ranked by eCPM which is calculated by the product of the predicted click-through-rate (pCTR) and bid of ad, and the ad with the maximum eCPM wins and its payment is the eCPM of the next rank ad divided by its own pCTR. Evidently, the payment for the winning ad depends on the bid of the next ranked ad, not its own bid. Meanwhile, the organic items list are sorted by the estimated GMV. Subsequently, the ad allocation algorithm dynamically inserts the auction-winning ad into the ordered organic items list, where the winning ad's payment is used for positions decision \cite{liao2021cross}, aiming to maximize revenue and GMV. From the perspective of ad mechanism design, this method of segregating the ad auction and allocation into two distinct stages face two problems as follows: 

\begin{itemize}[leftmargin=*]
    \item \textbf{Ad auction does not take externalities into account}. Ad externalities generally refer to the impact of its display position and context on its CTR \cite{ghosh2008externalities}. Traditional auctions, including the GSP, often do not consider externalities and obtain stable results based on the separated CTR assumption \cite{varian2007position}. However, in practice, the CTR of ad surrounded by organic items is impacted by its actual display position and context. Externalities result in complicated strategic competition among advertisers, precluding stable outcomes and social welfare optimization \cite{qin2015sponsored}. Hence, when considering the externalities, traditional auctions are not suitable.
    \item \textbf{Ad allocation, using the winning ad's payment to determine display position dynamically is not IC for the ad}. For the winning ad, even if its bid is increased, it does not secure a better position, thereby losing the opportunity to gain greater utility. This is because in the aforementioned ad auction, even if the winning ad increases bid, its payment remains the same, and thus the ad allocation outcomes do not change. Furthermore, using the bid of the winning ad to determine the display position is also inappropriate. Similarly, without any change in payment, the winning ad is motivated to increase its bid until it secures the best position and maximum utility. However, at this juncture, the GMV loss of the platform may exceed the increase in revenue.
\end{itemize}

In order to solve these problems, designing a mechanism that integrates ad auction and allocation in feed should be considered. This mechanism needs to consider externalities and satisfy the economic properties of IC and IR \cite{myerson1981optimal, li2023learning}. IC means that when the advertiser truthfully reports his value, his utility is maximized. IR means that the advertiser's payment will not be higher than his bid. Under the externalities, The ad's CTR varies with its position and context, which makes it not easy to calculate the social welfare of the mechanism \cite{ li2023learning}. Consequently, designing an externality-aware mechanism that satisfies both IC and IR is very difficult. VCG \cite{edelman2007internet}, another classic mechanism, provides ideas for solving these problems. But VCG's expected revenue is lower than that of GSP \cite{edelman2007internet, varian2007position}, which is not acceptable for the platform. This mechanism also needs to maximize the platform's revenue and GMV. 

Therefore, this paper proposes a deep automated \textbf{M}echanism for \textbf{I}ntegrating ad \textbf{A}uction and \textbf{A}llocation in feed, referred to as \textbf{MIAA}, which satisfies both IC and IR in the presence of externalities and maximizes revenue and GMV. Inspired by affine maximizer auction(AMA) \cite{sandholm2015automated} based VCG, MIAA is designed to include the following three parts. Firstly, MIAA takes candidate ads and the ordered list of organic items as input. For each candidate ad, several candidate allocations are generated by inserting the ad in different positions of the ordered organic items list. Secondly, for each candidate allocation, a list-wise prediction model takes the entire allocation as input and outputs the predicted results for each ad and organic item to model the global externalities. Finally, an automated auction mechanism is executed to select the optimal allocation to display from these candidate allocations. The mechanism parameters modeled by deep neural networks are learned in an end-to-end way to maximize platform revenue and GMV. Obviously, MIAA simultaneously obtains the ranking, payment, and display position of ad. The contributions of our work are summarized as follows:
\begin{itemize}[leftmargin=*]
    \item \textbf{A deep automated mechanism for integrating ad auction and allocation}. This mechanism simultaneously decides the ranking, payment, and display position of the ad. To the best of our knowledge, the proposed mechanism is the first to consider addressing the externalities and IC issues caused by the separation of ad auction and allocation.
    \item \textbf{Detailed industrial and practical experience}. We successfully deploy MIAA on the Meituan retail feed and obtain significant improvements in both platform revenue and GMV.
 \end{itemize}

 \section{related work}
 Previous work largely focused on studying optimization of ad auction and allocation separately. The main points of these work are summarized as follows::
 \begin{itemize}[leftmargin=*]
     \item \textbf{Externality-aware ad auction design}. The design of ad auction with externalities has always been a hot topic in both academic and industrial research. The CTR model considering the impact of ad externalities has evolved from local model to global model. Ghosh and Mahdian \cite{ghosh2008externalities} initially studied the externalities of user click interactions between advertisers. They used a choice model to describe user click behaviors and adopted a VCG-style payment scheme to guarantee IC and IR. User browsing model \cite{dupret2008user} and cascade models \cite{kempe2008cascade, fotakis2011externalities, farina2016ad}, built based on user browsing behavior from top to bottom of the page, take the local impact of externalities into consideration. And then ad auction with user browsing model or cascade models are designed based on GSP or VCG. DNA \cite{liu2021neural} and DPIN \cite{huang2021deep} respectively modeled the local externalities on candidate ad sets and ad positions. EXTR \cite{chen2022extr} modeled the global externalities, but did not consider the subsequent auction mechanism design. The global externalities, such as the interaction between ads and organic items, were considered in \cite{liao2022nma, li2023learning, gatti2012truthful}. Similarly, this paper directly models the allocation list mixed with organic items and an ad, taking the global externalities into account.
    \item \textbf{Ad allocation}. The platform initially assigned fixed positions to organic items and ads. Dynamic ad allocation strategies to optimize the whole-page performance gains growing attention. These strategies can be divided into two categories. In the first category, the dynamic ad allocation strategies, proposed by \citet{yan2020ads, zhang2018whole, zhao2019deep, xie2021hierarchical, xu2023multi}, do not take into account the relative ranking of candidate ads in auction, which dose not satisfy the IC and IR restrictions. In the second category, the strategies maintain the relative ranking of ads and determine the optimal number and positions of ads in the mixed list \cite{wang2022hybrid,wang2022learning,shi2023mddl,liao2022deep}. Liao et al.\cite{liao2021cross} propose a DQN architecture to incorporate the arrangement signal into the allocation model without modifying the relative ranking of advertisements. Similarly, Chen et al.\cite{chen2022hierarchically} propose hierarchically constrained adaptive ad exposure that possesses the desirable game-theoretical properties and computational efficiency.
\end{itemize}

Recently, Li et al. \cite{li2023optimally} prospectively analyze that this separated method in search scenarios is detrimental to the platform user experience. They propose an integrated model to describe the mixed arrangements of organic items and  ads. Based on this model, they design a truthful optimal mechanism with the separated CTR assumption and publicly known distribution of advertising value. Furthermore, this paper breaks the separated CTR assumption and aims to design deep automated mechanism for integrating ad auction and allocation in feed, which satisfies both IC and IR in the presence of externalities and maximizes revenue and GMV.

\section{Preliminaries}

\subsection{Setting for Ad Auction and Allocation}
This paper considers the mechanism design that integrates pay-per-click ad auction and allocation in feed. For each user page view request, there are $m$ available positions for placing ads and organic items. There are $n$ advertisers competing for one ad position and other $m-1$ positions are filled with organic items. Due to the limitations of platform business division, we assume that these organic items have been sorted according to the estimated GMV. In the subsequent process, the relative ranking of organic items will not be modified, but their CTR and GMV will be re-predicted. The main notations in this paper are summarized in Table \ref{table:notation}.

Each advertiser $i \in [n]$ has a private click value $v_i \in \mathcal{B} \subseteq{R}^{+}$ on his ad and reports a click bid $b_i \in \mathcal{B} \subseteq{R}^{+}$ for auction. We assume that the click value and bid for the organic items are 0. Let $\textbf{v}=(v_1,v_2,\dots,v_n) \in \mathcal{B}^{n}$ and $\textbf{b}=(b_1,b_2,\dots,b_n) \in \mathcal{B}^{n} $ be the value profile and the bid profile of all ads. We use $\textbf{v}_{-i}$ and $\textbf{b}_{-i}$ to represent the value profile and bid profile of all ads except ad $i$. An ad allocation can be represented by $\textbf{a}=a(i,j)$, where $a(i,j)$ means that the ad $i$ is inserted into the $j$-th position. Let $j=\sigma(i)$ represent the position index of ad $i$ in an allocation. The set of all possible allocations is denoted by $\mathcal{A}=\{a(i, j): \forall i \in [n], j \in [m]\}$. 


Formally, as shown in the Figure \ref{fig:fig_intro}, one of the allocations is an ordered list mixed with several organic items and an ad. When modeling the impact of the global externalities, the CTR prediction needs to take various aspects into consideration, such as attributes of the ad itself, the position and context of the ad, and the attributes of surrounding organic items \cite{li2023learning}. We denote the pCTR of the $j$-th item in the allocation $\textbf{a}$ by $q_{j}(\textbf{a})$. Let $g_j(\textbf{a})$ be the estimated merchandise volume per click of the $j$-th item in the allocation. For convenience, we use $e_{j} \in \mathcal{E} \subseteq{R}$ to represent the externalities such as CTR and GMV for the $j$-th item in the allocation.

\begin{table}[tb]
  \vspace{0.05in}
  \centering
  \caption{Main notations.}
  \setlength{\tabcolsep}{3pt}
  \begin{tabular}{c|c}
    \hline
    Notation & Meaning \\
    \hline
    $ad$ & Ad item  \\
    $oi$ & Organic item  \\
    $m$ & Number of available positions  \\
    $n$ & Number of ads  \\
    $v_{i} \in \mathcal{B}$ & Real click value of $a_i$ \\
    $\textbf{v}=(v_1,\dots,v_n)$ & Values profile of ads \\
    $\textbf{v}_{-i}$ & Values profile of ads except ad $i$ \\
    $b_{i} \in \mathcal{B}$ & Submitted click bid of $a_i$ \\
    $\textbf{b}=(b_1,\dots,b_n)$ & Bids profile of ads  \\
    $\textbf{b}_{-i}$ & Bids profile of ads except ad $i$ \\
    $\textbf{a} \in \mathcal{A}$ & An ad allocation \\
    $\mathcal{A}_{-i}$ & all the ad allocations except ad $i$ \\
    $\textbf{a}_{-i} \in \mathcal{A}_{-i}$ & An ad allocation except ad $i$ \\
    $\sigma(i)$ & Position index of ad $i$ in an allocation \\
    $q_{j}(\textbf{a})$ & pCTR of the $j$-th item in $\textbf{a}$ \\
    $g_j$ & Estimate GMV per click of the $j$-th item \\
    $e_j(\textbf{a})$ & Externality of the $j$-th item in $\textbf{a}$ \\
    $\textbf{e}(\textbf{a})$ & Externality of the items in $\textbf{a}$ \\
    $\mathcal{M} \langle \mathcal{R},\mathcal{P} \rangle$ & Ad auction and allocation mechanism \\
    $\mathcal{R}$ & Allocation rule \\
    $\mathcal{P}$ & Pricing rule rule \\
    $\text{Rev}(\textbf{b};\textbf{e})$ & Platform expected revenue \\
    $\text{Gmv}(\textbf{b};\textbf{e})$ & Platform expected GMV \\
    \hline
  \end{tabular}
  \vspace{0.05in}
  \label{table:notation}
  \vspace{-0.15in}
\end{table}

\subsection{Problem Formulation}
Given bids from advertisers and the ordered organic items list, we represent an ad auction and allocation mechanism $\mathcal{M} \langle \mathcal{R},\mathcal{P} \rangle$, where:
\begin{itemize}
\item $\mathcal{R}(\textbf{b};\textbf{e}): \mathcal{B}^{n} \times \mathcal{E}^{m} \to \mathcal{A}$ is the ad allocation rule, which is used to select one winning ad from the $n$ advertisers and insert it at the certain position in the organic items list.
\item $\mathcal{P}(\textbf{b};\textbf{e})$ is the payment rule and $p_i(\textbf{b};\textbf{e})$ is the pay-per-click price of advertiser $i$.
 \end{itemize}

For a mechanism $\mathcal{M} \langle \mathcal{R},\mathcal{P} \rangle$, the expected utility of advertiser $i$ is 
\begin{equation}
u_{i}^{\mathcal{M}}(v_i;\textbf{b};\textbf{e})=(v_i - p_i(\textbf{b};\textbf{e})) \times q_{\sigma(i)}(\mathcal{R}(\textbf{b};\textbf{e})),
\end{equation}
and the expected revenue and GMV of the platform is
\begin{equation}
\begin{aligned}
\label{eq:RevAndGMV}
\text{Rev}^{\mathcal{M}}(\textbf{b};\textbf{e})=& \sum_{i=1}^{n} p_i(\textbf{b};\textbf{e})) \times q_{\sigma(i)}(\mathcal{R}(\textbf{b};\textbf{e})), \\
\text{Gmv}^{\mathcal{M}}(\textbf{b};\textbf{e})=& \sum_{j=1}^{m} g_j(\mathcal{R}(\textbf{b};\textbf{e})) \times q_{j}(\mathcal{R}(\textbf{b};\textbf{e})).
\end{aligned}
\end{equation}

For the design of ad auction mechanisms, IC and IR are standard economic constraints that must be considered \cite{li2023learning}. An auction mechanism $\mathcal{M} \langle \mathcal{R},\mathcal{P} \rangle$ is IC, if for each advertiser truthfully reports his bid $b_i = v_i$ and then his utility is maximized. Formally, for any $\textbf{e}$, for each $i$, it holds that
\begin{equation}
u_{i}(v_i;v_i, \textbf{b}_{-1};\textbf{e}) \ge u_{i}(v_i;b_i, \textbf{b}_{-1};\textbf{e}), \forall b_i \in \mathcal{B},
\end{equation}
and an auction mechanism $\mathcal{M} \langle \mathcal{R},\mathcal{P} \rangle$ is IR, if for each advertiser would not be charged more than his bid for the allocation. Formally, for any $\textbf{e}$, for each $i$, it holds that
\begin{equation}
p_i(\textbf{b};\textbf{e}) \le b_i.
\end{equation}

We aim to design a mechanisms $\mathcal{M} \langle \mathcal{R},\mathcal{P} \rangle$ which satisfies both IC and IR in the presence of externalities and maximizes revenue and GMV of the platform, as follows:
\begin{equation}
  \begin{aligned}
  \maximize_{\mathcal{M}} \quad & \mathbb{E}_{\mathbf{a} \in \mathcal{A}} [\text{Rev}(\textbf{b};\textbf{e}) + \alpha \text{Gmv}(\textbf{b};\textbf{e})],\\
  \textrm{s.t.} \quad 
  & \textit{IC and IR constraint,}
  \end{aligned}
  \label{eq:problem}
\end{equation}
where $\alpha$ is a weight coefficient set by the platform to balance revenue and GMV. 

\subsection{Automated Mechanism Design}
Formally, the integration of ad auction and allocation can be seen as a combination auction (CA) of ads and organic items, where the private value and bid of organic items on click can be assumed to be 0. Sandholm and Likhodedov \cite{sandholm2015automated} proposed an automated mechanism based VCG named AMA for revenue-maximizing combinatorial auctions. The AMA is defined as follows.

Each bidder $j$ submits a valuation function $v_j$. The allocation, $\textbf{a}^{*}$, is computed to maximize
\begin{equation}
\text{SW}^{\mu}_{\lambda}(\textbf{a}) = \sum_{j=1}^{m} {\mu}_{j}v_j(\textbf{a}) + {\lambda}(\textbf{a}),
\end{equation}
where ${\mu}_{j}$ is a positive number that is related to the value distribution of the bidder $j$ but not related to the allocation, and meanwhile ${\lambda}(\textbf{a})$ is an arbitrary function of the allocation. The payments are
\begin{equation}
p_{j}(\textbf{a}^{*}) = \frac{1}{{\mu}_{j}}\left[\text{SW}^{\mu}_{\lambda}(\textbf{a}^{*}_{-j}) - \sum_{i \ne j }{\mu}_{i}v_i(\textbf{a}^{*}) - {\lambda}(\textbf{a}^{*})\right],
\end{equation} 
where 
\begin{equation}
\textbf{a}^{*}_{-j} = \maximize_{\textbf{a} \in \mathcal{A}_{-j}}\text{SW}^{\mu}_{\lambda}(\textbf{a}).
\end{equation} 
$\textbf{a}^{*}_{-j}$ is the best allocation where bidder $j$ is not present.

AMAs are a family of mechanisms, parameterized by the vectors ${\mu}$ and ${\lambda}$. VCG is the special case where for all bidders $j$ and any allocation, ${\mu}_{j} = 1$ and ${\lambda}(\cdot) = 0$. The ${\mu}_{j}$ and ${\lambda}(\cdot)$ are solved using automated search theory to maximize revenue \cite{sandholm2015automated}. Roberts et al. \cite{roberts1979characterization} and Lavi et al.\cite{lavi2003towards} has proven that only AMAs are IC and IR for all CA settings. Hence, inspired by AMA, this paper extends its application to integrate ad auction and allocation. Unlike search theory, in this paper, ${\mu}_{j}$ and ${\lambda}(\cdot)$ are modeled as deep neural networks and trained by end-to-end learning methods.

\section{Methodology}
\begin{figure*}[tb]
  \centering
  \includegraphics[width=0.95\textwidth]{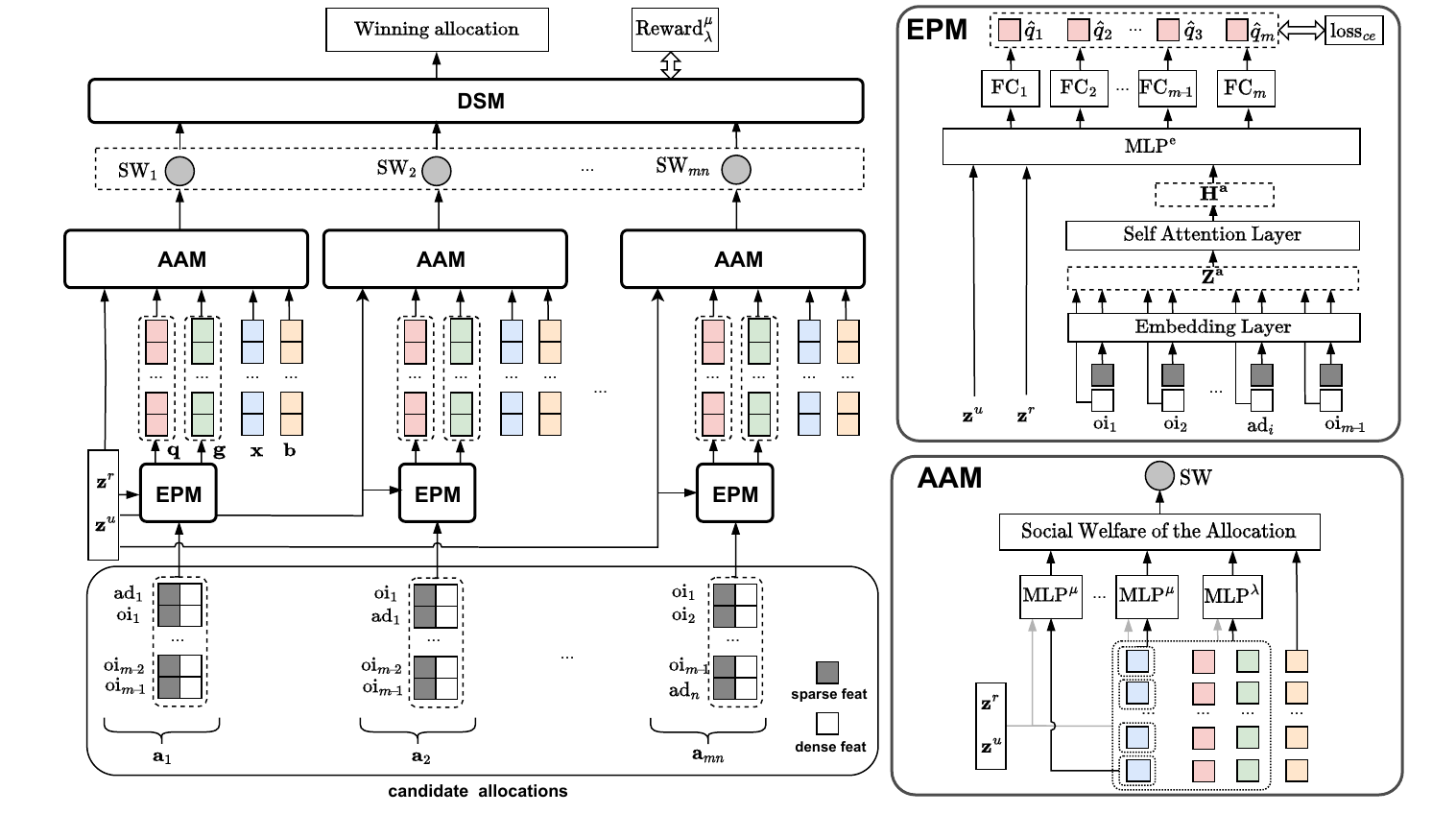}
  \caption{
    The architecture of MIAA. MIAA consists of three modules: Externality-aware Prediction Module (EPM), Automated Auction Module (AAM) and Differentiable Sorting Module (DSM).
  }
  \label{fig:iaa}
\end{figure*}

In this section, we present the details of deep automated mechanism for integrating ad auction and allocation in feed, namely MIAA, which satisfies both IC and IR in the presence of externalities and maximizes revenue and GMV. As shown in Figure \ref{fig:iaa}, MIAA takes candidate ads and the ordered organic items list as input, and outputs the optimal allocation by three modules. The three modules of MIAA are the Externality-aware Prediction Module (EPM), the Automated Auction Module (AAM) and the Differentiable Sorting Module (DSM). More specifically, EPM takes an allocation as inputs to model the impact of the global externalities and outputs the predicted results for each ad and organic item. AAM uses two deep neural networks to model the two mechanism parameters ${\mu}_{j}$ and ${\lambda}(\cdot)$ with the aim of improving the expressive ability of this automated mechanism while guaranteeing IC and IR. DSM uses multi classification model softmax to conduct a continuous relaxation of sorting operator in the mechanism, and outputs a vector representing the winning probability for each candidate allocation. And then the expected revenue and GMV of the platform can be differentiably calculated and optimized by an end-to-end learning way. Next we will introduce each module separately.

\subsection{Externality-aware Prediction Module}
In most traditional auction mechanisms, the position and contextual information of ads can only be known after the auction, so the CTR prediction model cannot obtain them in advance. However, auctions rely on the pCTR by the prediction model. In order to solve this interdependence problem and obtain stable allocation result, the mechanism design is based on the separated CTR assumption, that is, the final CTR of an ad is equal to the product of the CTR of its own content and the CTR of its position, ignoring the influence of contextual items. Therefore, under this assumption, the point-wise model is often used to predict CTR, which do not consider the impact of ads display positions and interaction between ads and organic items \cite{varian2007position}. Several models \cite{chen2022extr, liu2021neural, huang2021deep} capturing local externalities have been proposed successively, focusing solely on the display position of ad or the local context of ad. To this end, the paper uses a list-wise prediction module to model global externalities explicitly, outputting the more accurate pCTR for each item.

Firstly, for each candidate ad, several candidate allocations are generated by inserting the ad in different positions of the ordered organic items list. This step corresponds to a time complexity of $O(m \times n)$. In actual production processes, the value of $m$ is typically not greater than 5, which is acceptable in terms of performance for the platform.

Then, EPM adopts a parameter-sharing structure for all the candidate allocations. Here we take one allocation $\textbf{a}$ as an example to illustrate. As shown in Figure \ref{fig:iaa}, EPM takes the allocation $\textbf{a}$ and two types of public information (i.e., request information, user profile in current request) as input and outputs the pCTR of each item in the allocation. We first use embedding layers to extract the embeddings from raw sparse features and then concatenate dense features with the embeddings to form $d$-dimensional feature vector $\textbf{z}_j \in \mathbb{R}^{d}$ for the $j$-th item. The feature matrix of the allocation $\textbf{a}$ is represented as $\textbf{Z}^{\textbf{a}} \in \mathbb{R}^{m \times d}$, while the feature vector of request information, user profile are respectively represented by $\textbf{z}^u$, and $\textbf{z}^r$. 

Next, We use a Self Attention Unit \cite{vaswani2017attention} to model the interaction between the ad and the organic items in the candidate allocation:
\begin{equation}
\textbf{H}^{\textbf{a}} = \text{SelfAtt} \left(\textbf{Q}^{\textbf{a}},\textbf{K}^{\textbf{a}},\textbf{V}^{\textbf{a}}\right) = \text{softmax}\left(\frac{\textbf{Q}^{\textbf{a}}(\textbf{K}^{\textbf{a}})^{\top}}{\sqrt{d}}\right)\textbf{V}^{\textbf{a}},
\end{equation}
where $\textbf{Q}^{\textbf{a}}$, $\textbf{K}^{\textbf{a}}$, $\textbf{V}^{\textbf{a}}$ represent query, key, and value, respectively. Here query, key, and value are transformed linearly from feature information of the allocation $\textbf{a}$ , as follows:
\begin{equation}
\textbf{Q}^{\textbf{a}}=\textbf{Z}^{\textbf{a}}\textbf{W}^Q,\textbf{K}^{\textbf{a}}=\textbf{Z}^{\textbf{a}}\textbf{W}^K,\textbf{V}^{\textbf{a}}=\textbf{Z}^{\textbf{a}}\textbf{W}^V.
\end{equation}

Finally, we reshape $\textbf{H}^{\textbf{a}}$ as one vector $\textbf{h}^{\textbf{a}} \in \mathbb{R}^{md} $ and concatenate $\textbf{z}^u, \textbf{z}^r$ and put them into a Multi-Layer Perceptron (MLP) to model the global externalities:
\begin{equation}
 \begin{aligned}
 \hat{q}_{j} = q_{j}(\textbf{a})=& \text{Sigmoid} \left(\text{FC}_{j}\left(\text{MLP}^{e}(\textbf{h}^{\textbf{a}}|\textbf{z}^u|\textbf{z}^r)\right) \right), \forall j \in [m],
 \end{aligned}
\end{equation}
where $q_{j}(\textbf{a})$ denotes the pCTR of the $j$-th item in the allocation. In EPM, 
the cross entropy loss is calculated by the real click behavior of each item to train this list-wise model:
\begin{equation}
\text{Loss}_{ce}= -\sum_{j=1}^{m} \left(y_{j} \text{log}(q_{j}(\textbf{a}) + (1-y_{j})\text{log}(1-q_{j}(\textbf{a}) \right),
\end{equation}
where $y_{j} \in \{0, 1\}$ represents whether user clicks the $j$-th item in the allocation or not. 

EPM takes a candidate allocation as input and outputs the pCTR for each position in the allocation. It is built and trained independently, without coupling to downstream modules. EPM is a general framework that can easily be extended to multiple targets prediction such as conversion rate (CVR) and GMV. The $g_{j}(\textbf{a})$ used in the following text can be predicted simultaneously in EPM, without further elaboration. Compared with point-wise models, list-wise models capture the global externalities and output more accurate results, which can help the subsequent modules achieve better performance.

\subsection{Automated Auction Module}
This module selects the optimal allocation from all possible candidate allocations, which simultaneously decides the ranking, payment, and display position of the ads. Extending AMA into a deep automated mechanism, $\mu$ and $\lambda(\cdot)$ are modeled as deep neural networks $\mu$-network and $\lambda$-network to improve the expressive ability of ranking formula while guaranteeing IC and IR properties.

Firstly, $\mu$ represents the strength of item bidding capability and is independent of the allocations \cite{sandholm2015automated}. Thus, more specifically, the value distribution information of ads, the features of request information, user profile and items are the main input features of the $\mu$-network. Let $\textbf{x}_j$ represent the value distribution information vector of the $j$-th item in the allocation. The $\mu$-network can be formally expressed as:
\begin{equation}
f^{\mu}_{j} = \text{Sigmoid} \left(\text{MLP}^{\mu}(\textbf{x}_j|\textbf{z}^u|\textbf{z}^r) \right), \forall j \in [m],
\end{equation}
where the sigmoid function ensures a positive output. Each item has a $f^{\mu}_{j}$ to represent its competitive ability.

Then, $\lambda$ is an arbitrary function of the allocation $\textbf{a}$. The features of each item in the allocation $\textbf{a}$ are concatenated to represent the entire allocation, and the $\lambda$-network for the allocation $\textbf{a}$ is:
\begin{equation}
f^{\lambda}(\textbf{a}) = \text{MLP}^{\lambda} \left(\textbf{o}_1|\textbf{o}_2|\dots|\textbf{o}_m|\textbf{z}^u|\textbf{z}^r\right),
\end{equation}
where $\textbf{o}_j= \textbf{x}_j|q_j(\textbf{a})|g_j(\textbf{a})$.

Finally, the social welfare of the allocation $\textbf{a}$ is calculated as:
\begin{equation}
\text{SW}^{\mu}_{\lambda}(\textbf{a}) = \sum_{j=1}^{m} f^{\mu}_{j} \times \text{eCPM}_{j}(\textbf{a}) + f^{\lambda}(\textbf{a}),
\end{equation}
where $\text{eCPM}_{j}(\textbf{a})= b_j \times q_{j}(\textbf{a})$ is the $j$-th item valuation function on the allocation $\textbf{a}$. 

The optimal allocation $\textbf{a}^{*}$ is computed to maximize $\text{SW}^{\mu}_{\lambda}(\cdot)$. The payment on click of the ad $i$ in the $\sigma(i)$-th position of $\textbf{a}^{*}$ are
\begin{equation}
\label{eq:111}
p_{i}(\textbf{a}^{*}) = \frac{1}{f^{\mu}_{\sigma(i)}(\textbf{a}^{*})}\left[\text{SW}^{\mu}_{\lambda}(\textbf{a}^{*}_{-i}) - \text{SW}^{\mu}_{\lambda}(\textbf{a}^{*})_{-i}\right] \times \frac{1}{q_{\sigma(i)}(\textbf{a}^{*})},
\end{equation} 
where $\textbf{a}^{*}_{-i}$ is the optimal allocation where ad $i$ is not present and $\text{SW}^{\mu}_{\lambda}(\textbf{a}^{*})_{-i} = \sum_{j \ne \sigma(i) }f^{\mu}_{j} \cdot  b_j \cdot q_{j}(\textbf{a}^{*}) + f^{\lambda}(\textbf{a}^{*})$. Because there is only one ad in the allocation, and the bid of each organic item is 0, so $\text{SW}^{\mu}_{\lambda}(\textbf{a}^{*})_{-i} = f^{\lambda}(\textbf{a}^{*})$.

Obviously, the process maintains the IC and IR properties of AMA. The $\mu$-network and $\lambda$-network are trained by end-to-end learning in the next module DSM.

\subsection{Differentiable Sorting Module}
The search theory used in AMA \cite{sandholm2015automated} to solve $\mu$ and $\lambda$ parameters is inefficient. This paper aims to improve the efficiency and effectiveness of solving this mechanism problem in the end-to-end learning way. However, We are faced with two challenges, the first one is the non-differentiability of the ranking process in auctions, and the second one is the lack of user real behavior feedback. 

In AAM, the sorting operation to obtain the optimal allocation leads to a non-differentiable process for the entire procedure. Inspired by the differentiable sorting engine proposed by Liu et al. \cite{liu2021neural}, we use multi classification model softmax to conduct a continuous relaxation of sorting operator in the mechanism. Given the allocations set $\textbf{SW}=[\text{SW}^{\mu}_{\lambda}(\textbf{a}_{1}), \dots, \text{SW}^{\mu}_{\lambda}(\textbf{a}_{mn})]$, 
the vector $\textbf{Pr}$ of the allocations is mapped by the softmax function:
\begin{equation}
\textbf{Pr} = \text{softmax}(\frac{\textbf{SW}}{\tau})=[Pr(\textbf{a}_{1}), Pr(\textbf{a}_{2}), \dots, Pr(\textbf{a}_{mn})],
\end{equation}
where $\tau$ is a temperature parameter. Intuitively, $\textbf{Pr}$ can be interpreted as the winning probabilities on all allocations. 

We use the pCTR and preditcted GMV (pGMV) indicators to simulate user behavior, and calculate the performance metrics of the allocations, which are then used for feedback training. Hence, the expected revenue and GMV of the allocations are calculated by Eq.(\ref{eq:RevAndGMV}) as:
\begin{equation}
  \begin{aligned}
\textbf{Rev} =& [\text{Rev}(\textbf{a}_{1}), \text{Rev}(\textbf{a}_{2}), \dots, \text{Rev}(\textbf{a}_{mn})], \\
\textbf{Gmv} =& [\text{Gmv}(\textbf{a}_{1}), \text{Gmv}(\textbf{a}_{2}), \dots, \text{Gmv}(\textbf{a}_{mn})].
  \end{aligned}
\end{equation}
Specifically, only the winning allocation's revenue is greater than 0, while the non-winning allocations' revenues are 0. Finally, The overall optimization goal is to maximize $\text{Reward}^{\mu}_{\lambda}$, i.e., minimize 
\begin{equation}
\text{Loss} = -\text{Reward}^{\mu}_{\lambda} = -\sum_{k =1}^{mn} Pr(\textbf{a}_{k}) \left( \text{Rev}(\textbf{a}_{k}) + \alpha \text{Gmv}(\textbf{a}_{k}) \right).
\end{equation}
This $\text{Loss}$ provides direct feedback for training $\mu$-network and $\lambda$-network, which is a partial differentiable method. This training process is highly dependent on the accuracy of the pCTR and pGMV, which may lead to inconsistencies between the offline and online effects of this mechanism. Therefore, it is necessary to improve the prediction performance of pCTR and pGMV in EPM and perform list-wise calibration on them \cite{deng2021calibrating}.

\subsection{Training and Online Serving}
The data (e.g., the candidate ads set, the organic items list, and the real display ordered list and so on) in each request should be recorded for the training of MIAA. The training of $\mu$-network and $\lambda$-network in AAM relies on the prediction output of the list-wise model in EPM. Hence, the list-wise model in EPM first needs to be trained separately, using real display ordered list data. Then, for each request record, all the possible allocations are generated based on candidate ads and the organic items list. Next, the trained list-wise model in EPM provides the prediction results of pCTR and pGMV for each allocation, which will be used for subsequent training of $\mu$-network and $\lambda$-network in AAM. This training process cannot obtain real user behavior feedback, so it relies on prediction data for evaluating model performance.

The trained models in EPM and AAM are deployed online simultaneously and online serving is provided according to the following process:
\begin{itemize}[leftmargin=*]
\item \textbf{Step1}. Generate all the possible allocations.
\item \textbf{Step2}. Predict the pCTR and pGMV by EPM for each allocation.
\item \textbf{Step3}. Select the optimal allocation by AAM from all the possible allocations.
\item \textbf{Step4}. Return the winnning ad and its position.
\end{itemize}

\begin{table}[tb]
  \caption{Statistics of the datasets. Avg items represents the average number of items per request.\ \ }
  \renewcommand\arraystretch{1.2}
  \centering
  \setlength{\tabcolsep}{2mm}{
  \begin{tabular}{c|cccc}
    \hline
  Dataset & Requests  &  Users  & Items  & Avg items \\
  \hline
  \hline
  Avito & 112,159,463  & 4,284,824   &  36,893,299 & 3.64 \\
  Meituan & 232,171,262  & 65,459,719  & 3,889,579 & 4.74 \\
  \hline
  \end{tabular}
  }
  \label{dataset}
\end{table}

\section{Experiments}
In this section, we evaluate the effectiveness of the proposed mechanism MIAA with the aim of answering the following questions:
\begin{itemize}[leftmargin=*]
\item \textbf{Q1}: Compared with the point-wise pCTRs, how does our list-wise model perform in terms of CTR prediction?
\item \textbf{Q2}: Compared with widely used ad auction and allocation mechanisms in the industrial platform, how does our mechanism perform in terms of platform revenue and GMV?
\end{itemize}

We conduct extensive offline experiments on public and industrial datasets and online A/B tests on Meituan retail delivery platform.

 \begin{table}[bp]
  \caption{Experiment results about AUC and PCOC on Avito dataset. 'Pos' stands for the position in allocation.}
  \renewcommand\arraystretch{1.2}
  \centering
  \setlength{\tabcolsep}{2.3mm}{
  \begin{tabular}{c|c|cc}
  \hline
  Position & Method   & AUC  & PCOC  \\
  \hline
  \hline
  \multirow{2}{*}{All Pos}
  & EPM   & \textbf{0.7344 $\pm$ 0.0016} & \textbf{1.4875	$\pm$ 0.0165}\\
  & point-wise & 0.7308 $\pm$ 0.0011 &  1.6224	 $\pm$ 0.0605\\
  \hline
  \multirow{2}{*}{1st Pos}
  & EPM   & \textbf{0.7386 $\pm$ 0.0015} & \textbf{1.4952 $\pm$ 0.0262}\\
  & point-wise & 0.7349 $\pm$ 0.0013 &  1.6107	 $\pm$ 0.0608\\
  \hline
  \multirow{2}{*}{4th Pos}
  & EPM   & \textbf{0.7264 $\pm$ 	0.0018} & \textbf{1.4763 $\pm$ 	0.0146}\\
  & point-wise & 0.7225	 $\pm$ 0.0010 &  1.6394	 $\pm$ 0.0602\\
  \hline
  \end{tabular}
  }
  \label{tab:result_Avito}
\end{table}

\subsection{Experimental Settings}
\subsubsection{Dataset} In the offline experiments, we provide empirical evidence for the effectiveness of our mechanism on both public and industrial datasets. The statistics of the two datasets are summarized in Table \ref{dataset} and their detailed descriptions are as follows:
\begin{itemize}[leftmargin=*]
\item \textbf{Avito}\footnote{https://www.kaggle.com/competitions/avito-context-ad-clicks/data}. The public dataset is the Avito dataset used for the Avito context ad clicks competition of Kaggle. It is a random sample of previously selected users' searches on avito.ru during at least 26 consecutive days. In each search result page, only five items in position 1, 2, 6, 7, 8 are logged. And only the items (i.e., contextual ads) in the 1-st and 7-th positions are marked with whether the user clicked or not. For subsequent experiments, we simulate whether users click on items in the 2-nd and 6-th positions, based on the click behavior in the 1-st and 7-th positions. Let $\text{CTR}_{j}$ represent the CTR of position $j$. The $\text{CTR}_{1}$ and $\text{CTR}_{7}$ can be obtained through statistical analysis. Then, $\text{CTR}_{2}$ is simulated to follow a normal distribution $N(0.8\text{CTR}_{1}+0.2\text{CTR}_{7}, 0.1\text{CTR}_{1})$, while $\text{CTR}_{6}$ is simulated to follow a normal distribution $N(0.2\text{CTR}_{1}+0.8\text{CTR}_{7}, 0.1\text{CTR}_{7})$. The public information consists of search ID, user ID and search date, while the item information consists of ad ID, location ID, category ID and title. For each sample, we select the 1-st item and the 7-th item as candidate ads, and the remaining three items are treated as organic items. Therefore, each candidate allocation is composed of a candidate ad and these three organic items. Here we use the data from 20150425 to 20150517 as the training set and the data from 20150518 to 20150520 as the testing set to avoid data leakage.

\item \textbf{Meituan}. The industrial dataset is collected under GSP auction and fix positions on Meituan retail delivery platform during December 2023. Converted from each feed request, each sample contains all possible candidate allocations composed of candidate ads and organic items, as well as the final winning optimal allocation. Each candidate allocation is composed of a candidate ad and three organic items. In each candidate allocation, the information of each item consists of bid, sparse features (e.g., ID, category, brand and so on), dense features(e.g., historical CTR, sales volume, value distribution information and so on). Moreover, the point-wise pCTR in the production environment is recorded for subsequent performance comparison. According to the date of data collection, we divide the dataset into training and test sets with the proportion of 8:2.
\end{itemize}

 \begin{table}[bp]
  \caption{Experiment results about AUC and PCOC on Meituan dataset. 'Pos' stands for the position in allocation.}
  \renewcommand\arraystretch{1.2}
  \centering
  \setlength{\tabcolsep}{2.3mm}{
  \begin{tabular}{c|c|cc}
  \hline
  Position & Method    & AUC  & PCOC  \\
  \hline
  \hline
  \multirow{2}{*}{All Pos}
  & EPM   & \textbf{0.7077 $\pm$ 0.0005} & \textbf{0.9989 $\pm$ 0.0096}\\
  & point-wise & 0.6485 &  0.9075 \\
  \hline
  \multirow{2}{*}{1st Pos}
  & EPM   & \textbf{0.7191 $\pm$ 0.0004} & \textbf{0.9921 $\pm$ 0.0269}\\
  & point-wise & 0.6631 &  0.7246 \\
  \hline
  \multirow{2}{*}{2nd Pos}
  & EPM   & \textbf{0.7320 $\pm$ 0.0006} & \textbf{0.9924 $\pm$ 0.0502}\\
  & point-wise & 0.6988 &  0.9685 \\
  \hline
  \multirow{2}{*}{3rd Pos}
  & EPM   & \textbf{0.6812 $\pm$ 0.0009} & \textbf{1.0075 $\pm$ 0.0362}\\
  & point-wise & 0.6157 &  0.9804 \\
  \hline
  \multirow{2}{*}{4th Pos}
  & EPM   & \textbf{0.6617 $\pm$ 0.0003} & \textbf{0.9786 $\pm$ 0.0517}\\
  & point-wise & 0.6037 &  0.9002 \\
  \hline
  \end{tabular}
  }
  \label{tab:result_Meituan}
\end{table}

\begin{table*}[t]
  \renewcommand\arraystretch{1.2}
  \caption{Experiment results of the five methods on two datasets. Each result is presented in the form of mean $\pm$ standard deviation. The pRPM and pGPM represent the RPM and GPM values calculated based on pCTR and pGMV. The $\alpha$ is 0.5. Lift percentage means the improvement of the method in this line over the result of GSP and Fixed Position. Considering the confidentiality of commercial data, the experimental results on the Meituan dataset are presented based on GSP and Fixed Position.
}
  \setlength{\tabcolsep}{2.5mm}{
  \begin{tabular*}{0.95\textwidth}{c|c|ccc}
  \hline
  Dataset &  Model  & Revenue+$\alpha$GMV    & pRPM     & pGPM\\
  \hline 
  \hline
  \multirow{5}{*}{Avito}
  &  GSP and Fixed Position & 86,445.13 $\pm$ 12.30 & 2.9435 $\pm$ 0.0016 & 31.2695 $\pm$ 0.0053 \\
  &  GSP and Cross DQN      & 88,596.27 $\pm$ 5.43 (+2.49\%) & 2.9606 $\pm$ 0.0020 (+0.58\%) & \textbf{32.0757 $\pm$ 0.0014(+2.58\%)} \\
  &  Score-Weighted VCG  & 88,008.38 $\pm$ 7.20(+1.81\%) & \textbf{5.6909} $\pm$ \textbf{0.0006(+93.34\%)} & 30.4879 $\pm$ 0.0025(-2.50\%) \\
  &  IAS                & 88,613.62 $\pm$ 8.01(+2.51\%) &  5.004 $\pm$ 0.0040(+70.02\%) &  31.0604 $\pm$ 0.0018(-0.67\%) \\
  &  \textbf{MIAA}  & \textbf{91,069.80} $\pm$ \textbf{8.48(+5.35\%)} & 5.0732 $\pm$ 0.0007(+72.35\%) & 31.9563 $\pm$ 0.0032(+2.20\%)\\
  \hline 
  \hline 
 \multirow{5}{*}{Meituan} 
& GSP and Fixed Position  & 1.0  & 1.0  & 1.0 \\
& GSP and Cross DQN       &   1.0547 $\pm$ 0.0024(+5.47\%) &  1.0346 $\pm$ 0.0013(+3.46\%) &  1.0573  $\pm$ 0.0008(+5.73\%) \\
& Score-Weighted VCG   &   1.0123 $\pm$ 0.0031(+1.23\%) &   \textbf{1.1980}  $\pm$ \textbf{0.0022(+19.80\%)} &   0.9887 $\pm$ 0.0015(-1.13\%) \\
& IAS   &   1.0330 $\pm$ 0.0026(+3.30\%) &   1.1310 $\pm$ 0.0018(+13.10\%) &  1.0205   $\pm$ 0.0027(+2.5\%) \\
& \textbf{MIAA}   & \textbf{   1.1105  $\pm$ 0.0023 (+11.05\%)} &  1.1560 $\pm$ 0.0029(+15.60\%) & \textbf{1.1047} $\pm$ \textbf{0.0025(+10.47\%)}\\
  \hline
  \end{tabular*}}
  \label{tab:result}
\end{table*}

\begin{table*}[t]
  \renewcommand\arraystretch{1.2}
  \caption{Online A/B tests (2023.10.15-2023.12.15 two months) compared with GSP+Fixed Position and GSP+Cross DQN on promoting all performance metrics. The $\alpha$ is 0.3.}
  \setlength{\tabcolsep}{7.5mm}{
  \begin{tabular*}{0.95\textwidth}{c|ccc}
  \hline
  Model           & Revenue+$\alpha$GMV & RPM              & GPM  \\
  \hline 
  \hline
\% Improved Compared with GSP and Fixed Position   & +8.17\% & +11.25\% & +7.37\%  \\
\% Improved Compared with GSP and Cross DQN        & +4.87\% & +7.63\% & +4.15\%  \\
  \hline
  \end{tabular*}}
  \label{tab:result_online}
\end{table*}

\subsubsection{Evaluation Metrics}
We construct an offline simulation system based on the pCTR and pGMV to evaluate the effectiveness of MIAA. Each experiment is repeated 5 times with different random seeds and each result is presented in the form of mean $\pm$ standard. The following evaluation metrics are used in our offline experiments and online A/B tests.
\begin{itemize}[leftmargin=*]
\item \textbf{Click-Through Rate}. $\text{CTR} = \frac{\sum click}{\sum impression}$.
\item \textbf{AUC}. AUC stands for Area Under the Curve, which is often used to evaluate the performance of a model in machine learning. The closer the AUC value is to 1, the better the model performs.
\item \textbf{PCOC}. PCOC stands for Predicted CTR Over posterior CTR, which is a measure of prediction accuracy. If the PCOC value is close to 1, it indicates a high prediction accuracy.
\item \textbf{Revenue Per Mille}.  $\text{RPM} = \frac{\sum click \times payment}{\sum impression} \times 1000$.
\item \textbf{GMV Per Mille}. $\text{GPM} = \frac{\sum GMV}{\sum impression} \times 1000$.
\end{itemize}

\subsubsection{Baselines}
In terms of externalities modeling, we compare the pCTR of the proposed list-wise model with that of the point-wise model. In terms of platform revenue and GMV, we compare MIAA with the following four common auction and allocation mechanisms:
\begin{itemize}[leftmargin=*]
\item \textbf{GSP} \cite{varian2007position} and \textbf{Fixed Position}. This is a mechanism where the ad auction and allocation are separated into two stages. First, the GSP mechanism is used to select the winning ad, and then this ad is displayed in a fixed position.
\item \textbf{GSP} and \textbf{Cross DQN} \cite{liao2021cross}. The winning ad in GSP and the organic items form multiple combinations of candidate positions, which are evaluated and selected through cross DQN, finally determining the display position of the ad. 
\item \textbf{Score-Weighted VCG} \cite{li2023learning}. The Score-Weighted VCG framework decomposes the optimal auction design into two parts: designing a monotone score function and an matching-based allocation algorithm. But it does not consider the platform GMV and its incentive effect.
\item \textbf{IAS} \cite{li2023optimally}. IAS treats organic items as special ads with a bid of 0 and integrates both ads and organic items into the optimal auction to get their rankings.
\end{itemize}

\subsection{Offline Experiments}
\subsubsection{Performance Comparison of externalities modeling (Q1)} On the public Avito dataset, we need to construct point-wise model and list-wise model of EPM. The point-wise model is a deep neural network with multiple fully connected layers of $256 \times 128 \times 64 \times 32 \times 1$. For each request, The items in the 1-st, 2-nd, 6-th, and 7-th positions are extracted to form a list for constructing the list-wise model of EPM. As mentioned earlier, only the 1-st position and 4-th position in this list have real user click behavior data that can be used to evaluate the performance of these two models. The detailed experimental results on the Avito dataset are shown in Table \ref{tab:result_Avito}. For all positions in allocation, the list-wise pCTR in EPM improves the AUC by 0.0036 compared with the point-wise pCTR, and its PCOC is closer to 1. And at each position, the prediction results of the list-wise model in EPM is better than that of the point-wise model.

On the Meituan industrial dataset, the point-wise pCTR by DIN \cite{zhou2018DIN} model is already included, which is used in the production environment. Therefore, we only need to construct the list-wise model of EPM to predict CTR, and then compare it with the former.  The detailed experimental results on Meituan industrial dataset are shown in Table \ref{tab:result_Meituan}. From the results, the list-wise model in EPM significantly improves the AUC of all positions from 0.6485 of the point-wise model to 0.7077, and the PCOC of all positions is even closer to 1. Considering positional externalities, the list-wise model in EPM solves the problem of low predition at the first position, which has always existed in the point-wise model.

Obviously, considering the externalities (e.g., the display position and context of the ad), the list-wise model in EPM performs better on AUC and PCOC metrics than the point-wise models on both the public Avito dataset and Meituan industrial dataset.

\subsubsection{Performance Comparison of mechanisms (Q2)}
To validate the efficacy of the proposed mechanism in enhancing platform revenue and GMV, we implemente the GSP and Fixed Position, GSP and Cross DQN, Score-Weighted VCG and IAS on these two datasets for comparative analysis. On the public Avito dataset, we select the 1-st item and the 7-th item as candidate ads, and the remaining three items are treated as organic items.  Moreover, we provide a simulated bid for each ad and a simulated pGMV per click for each item. The bid of each ad is independently sampled from uniform distribution between 0.5 and 1.0. Meanwhile, the pGMV per click for organic items is independently sampled from uniform distribution between 3.5 and 6.0, and the pGMV per click for ads is independently sampled from uniform distribution between 2.0 and 4.0. On these two datasets, the fixed position is set as 2 in the GSP and Fixed Position baseline. Particularly, it should be pointed out that, as the offline simulation systems are unable to obtain actual user behavior data under different mechanisms, the assessment of these experiments' effectiveness is statistically derived from the pCTR and pGMV of users on items. Considering the confidentiality of commercial data, the experimental results on the Meituan dataset are presented based on GSP and Fixed Position. The detailed experiment results on public and industrial datasets are shown in Table \ref{tab:result}.  

From the experiment results, we can see that MIAA achieves obtains the highest revenue and GMV, outperforming all baseline mechanisms. Compared with GSP and Fixed Position, MIAA achieves significant improvements, as items with higher eCPM or GMV obtain better positions, and the payment of ads is increased.
MIAA, considering the incentive effect of organic items on ads, achieves higher pRPM compared with GSP and Cross DQN. Score-Weighted VCG achieves maximum pRPM without considering the loss of GMV. Compared with IAS, MIAA, which considers externalities, achieves higher pRPM and pGMV.

\subsection{Online Results}
We present the online experiments by deploying the proposed mechanism in Meituan retail delivery platform. In the production environment of feed, for each page request, the system returns an ordered list containing one ad and three organic items. There are two baselines, one is GSP and fixed position (that is, the ad is inserted at the second position), and the other one is GSP and Cross DQN, in which the ad is dynamically assigned to a certain position.

To demonstrate the performance of the proposed mechanism, we conduct online A/B tests with 5\% of whole production traffic from Oct 15, 2023 to Dec 15, 2023. In the real production environment, we cannot conduct A/B tests based on advertisers. The A/B tests using user-based traffic makes it difficult to evaluate the proposed mechanism's incentive effect on advertisers' bidding. Therefore, we only focus on the performance of revenue, GMV, RPM, and GPM in the experiments. In order to make fair and efficient comparisons between different baselines in production traffic, we will equalize the number of ad displays in the experimental group with that in the baselines. The experiment results in online A/B tests are shown in Table \ref{tab:result_online}. From the results, we can find that the proposed mechanism achieves the highest promotion for revenue, GMV, RPM and GPM compared with GSP and Fixed Position,  and GSP and Cross DQN.
\balance
\section{Conclusions}
This paper proposes a deep automated mechanism for integrating ad auction and allocation, that satisfies both IC and IR in the presence of externalities and maximizes revenue and GMV. Specifically, for all candidate allocations formed by the mixture of the ad and organic items, we use a list-wise model for global externalities modeling. Then, we use an automated mechanism based on deep neural networks to select the optimal allocation from the candidate allocation set in an end-to-end learning way. In this process, the ranking, payment, and display position of the ads are determined at the same time. We conduct extensive offline experiments on public and industrial datasets and online A/B tests on Meituan retail delivery platform. The offline experimental results and online A/B tests showed that the proposed mechanism significantly outperforms the other existing mechanism baselines. In future research, we aim to investigate efficient mechanisms design for integrating multi-slots ads auction and allocation.
  
\balance
\bibliographystyle{ACM-Reference-Format}
\newpage
\bibliography{MIAA}
\end{document}